\documentclass[10pt,conference,letterpaper]{IEEEtran}
\IEEEoverridecommandlockouts
\usepackage[a4paper, total={184mm,239mm}]{geometry}
\def\BibTeX{{\rm B\kern-.05em{\sc i\kern-.025em b}\kern-.08em
    T\kern-.1667em\lower.7ex\hbox{E}\kern-.125emX}}

\makeatletter
\newcommand\notsotiny{\@setfontsize\notsotiny\@vipt\@viipt}
\makeatother

\usepackage[utf8]{inputenc}
\usepackage[most]{tcolorbox}

\usepackage{booktabs}
\usepackage{rotating}
\usepackage{tablefootnote}
\usepackage{graphicx}
\usepackage{tabularx}
\usepackage{caption}
\usepackage{multirow}
\usepackage[noadjust]{cite} 
\usepackage{setspace}

\usepackage{placeins}
\usepackage{float}
\usepackage{amsmath}
\usepackage{amssymb} %decomment for arXiv
\usepackage{amsfonts} %
\usepackage{calligra} %
\usepackage{mathtools}
\usepackage{amsthm} % for neater definitions
\usepackage[hidelinks,bookmarks=false]{hyperref}
\usepackage{dblfloatfix} %allows double-column floats to be at bottom of page
\usepackage{array}
\usepackage{enumitem}

\usepackage{pgfplots}
\usepackage{pgfplotstable}
\usepgfplotslibrary{statistics}
\usepackage{cleveref}
\usepackage{cite}
\usepackage{amsmath,amssymb,amsfonts}
\usepackage{algorithmic}
\usepackage{graphicx}
\usepackage{textcomp}
\usepackage{xcolor}
\usepackage{multirow}
\usepackage{booktabs}
\makeatletter
\let\MYcaption\@makecaption
\makeatother
\usepackage[font=footnotesize]{subcaption}
\makeatletter
\let\@makecaption\MYcaption
\makeatother
\usepackage{xcolor, colortbl}

\usepackage{float}
\usepackage{textcomp}
\definecolor{linkcolor}{RGB}{219, 48, 122}

\usepackage{mwe}

\usepackage{algorithm}
\usepackage{algorithmic}

\usepackage{url}
\usepackage[most]{tcolorbox}
\usepackage{varwidth}
\tcbuselibrary{xparse}
\tcbuselibrary{skins}
\usepackage{pifont}
\usepackage{framed}
\usepackage{soul}
\usepackage{multirow}
\usepackage{array}
\newcolumntype{L}[1]{>{\raggedright\let\newline\\\arraybackslash\hspace{0pt}}m{#1}}
\newcolumntype{C}[1]{>{\centering\let\newline\\\arraybackslash\hspace{0pt}}m{#1}}
\newcolumntype{R}[1]{>{\raggedleft\let\newline\\\arraybackslash\hspace{0pt}}m{#1}}
\newcolumntype{H}{>{\collectcell\lstinline}l<{\endcollectcell}}
\usepackage{url}
\newcommand{\xmark}{\ding{55}}%

\input{preamble/preamblelistingsconf}
\newcommand{\ignore}[1]{{}}

\newcommand{\squishlist}{
	\begin{list}{$\bullet$}
		{ \setlength{\itemsep}{0pt}
			\setlength{\parsep}{1pt}
			\setlength{\topsep}{1pt}
			\setlength{\partopsep}{0pt}
			\setlength{\leftmargin}{0.9em}
			\setlength{\labelwidth}{1.5em}
			\setlength{\labelsep}{0.4em} } }
	\newcommand{\squishend}{
	\end{list}  }

\definecolor{graphFirst}{RGB}{2,136,209} % Light Blue 700
\definecolor{graphSecond}{RGB}{211,47,47} % Red 700
\definecolor{graphThird}{RGB}{245,124,0} % Orange 700
\definecolor{graphFourth}{RGB}{56,142,60} % Green 700
\definecolor{graphFifth}{RGB}{81,45,168} % Deep Purple 700
\definecolor{graphSixth}{RGB}{69,90,100} % Blue Grey 700
\definecolor{graphSeventh}{RGB}{251,192,45} % Yellow 700\definecolor{backgroundFirst}{RGB}{129,212,250} % Light Blue 200
\definecolor{backgroundSecond}{RGB}{239,154,154} % Red 200
\definecolor{backgroundThird}{RGB}{255,204,128} % Orange 200
\definecolor{backgroundFourth}{RGB}{165,214,167} % Green 200
\definecolor{backgroundFifth}{RGB}{179,157,219} % Deep Purple 200
\definecolor{backgroundSixth}{RGB}{176,190,197} % Blue Grey 200
\definecolor{backgroundSeventh}{RGB}{255,245,157} % Yellow 200

\usepackage{tabularx}

\author{%
    \IEEEauthorblockN{Jitendra Bhandari\IEEEauthorrefmark{1}, Johann Knechtel\IEEEauthorrefmark{2}, Ramesh Narayanaswamy\IEEEauthorrefmark{3},  Siddharth Garg\IEEEauthorrefmark{1}, and Ramesh Karri\IEEEauthorrefmark{1}
    }
    \IEEEauthorblockA{%
    \IEEEauthorrefmark{1}New York University, New York, USA 
    \IEEEauthorrefmark{2}New York University Abu Dhabi, UAE 
    \IEEEauthorrefmark{3}Synopsys, Sunnyvale, CA, USA}
}

\begin{document}

\title{LLM-Aided Testbench Generation and \\
Bug Detection for Finite-State Machines}
% \title{TB or Not TB? LLM-Aided Testbench (TB) Generation for Finite-State Machines}

%\author{}
% \author{Jitendra Bhandari, 
% Mohammed Nabeel,
% Likhitha Mankali, 
% Ozgur Sinanoglu,~\IEEEmembership{Senior Member,~IEEE},\\
% Ramesh Karri,~\IEEEmembership{Fellow,~IEEE}, 
% and Johann Knechtel,~\IEEEmembership{Member,~IEEE}
% \thanks{This work was supported in part by the NYU Center for Cybersecurity (CCS) and the NYUAD CCS.%
% 	%Ganesh Gore, Xifan Tang, Pierre-Emmanuel Gaillardon are supported by AFRL and DARPA under agreement number FA8650-18-2-7855, and Scott Temple, Pierre-Emmanuel Gaillardon are supported by AFRL and DARPA under agreement number FA8650-18-2-7849.
% 	\protect\\ 
% % \indent J. Bhandari and A. Khader Thalakkattu Moosa contributed equally to this work.\protect\\
% \indent J.~Bhandari, L.~Mankali, and R. Karri are with New York University, New York City, NY, 11201 USA. E-mail: \{jb7410, lm4344, rkarri\}@nyu.edu\protect\\
% \indent J.~Knechtel, M.~Nabeel, and O.~Sinanoglu are with New York University Abu Dhabi, UAE. E-mail: \{johann, mtn2, ozgursin\}@nyu.edu\protect
% }}

% \date{November 2019}
% \footnote{978-1-6654-3274-0/21/$31.00 ©2021 IEEE}
\maketitle

\IEEEpubidadjcol

\begin{abstract}
% This work explores how Large Language Models (LLMs) can be tailored for chip testing. A critical aspect of chip design is functional testing using testbenches to analyze the function and coverage of
% the Register-Transfer Level (RTL). 
% We investigate how to incorporate commercial-grade testing and simulation tool feedback to GPT-4 for improved testbench generation.
% Our case study yields promising results, demonstrating that iterative feedback from EDA tools can indeed improve test coverage. We extended our study to detect any incorrect implementation using our test coverage framework.
This work investigates the potential of tailoring Large Language Models (LLMs), specifically GPT3.5 and GPT4, for the domain of chip testing. A key aspect of chip design is functional testing, which relies on testbenches to evaluate the functionality and coverage of Register-Transfer Level (RTL) designs. We aim to enhance testbench generation by incorporating feedback from commercial-grade Electronic Design Automation (EDA) tools into LLMs. Through iterative feedback from these tools, we refine the testbenches to achieve improved test coverage. Our case studies present promising results, demonstrating that this approach can effectively enhance test coverage. By integrating EDA tool feedback, the generated testbenches become more accurate in identifying potential issues in the RTL design. Furthermore, we extended our study to use this enhanced test coverage framework for detecting bugs in the RTL implementations. 
\end{abstract}


\section{Introduction}
%%%% IC Design Cycle

Testbench generation is vital in the integrated circuit (IC) design cycle to ensure that chips are functional, meet design specifications, and are of high quality. Testbenches are designed to activate and monitor a wide range of design behaviors, ensuring that every part of a circuit is exercised during testing, leaving no function unchecked. This comprehensive testing approach not only ensures functional correctness but also provides valuable insights that can lead to design optimizations.
%These optimizations can enhance performance, reduce power consumption, and minimize the area of the chip, contributing to overall design efficiency.
Functional bugs tend to occur in a chip's control path, which is often implemented via Finite-State Machines (FSMs). These FSMs orchestrate the sequence of operations within the chip, making their correct function critical. Thus, achieving full FSM coverage is often a requirement for compliance with industry standards, particularly in sectors like aerospace and medical devices etc., where high reliability and safety are paramount~\cite{takahashi03}.

Large Language Models (LLMs)~\cite{openai2024gpt4} have revolutionized the way developers approach coding and testing, including the realm of hardware description languages (HDL). Recent work has demonstrated the utility of LLMs for various hardware-related tasks. For example, LLMs have been used for Verilog code generation~\cite{liu2023verilogeval,thakur2023autochip,lu2023rtllm,thakur2023verigen}, where they help in writing and optimizing HDL code. They have also been employed in generating assertions~\cite{kande2023llmassisted,fang2024assertllm}, which are critical for verifying the correctness of HDL designs. Additionally, LLMs have shown promise in scripting for electronic design automation (EDA) tools~\cite{wu2024chateda,liu2023chipnemo}, streamlining the design and verification process.
These methods either directly leverage foundation models like GPT, or fine-tune specialized code generation models on hardware datasets. Results from these studies suggest that LLMs can effectively understand Verilog code with its syntax, structure, and functional requirements. This understanding enables LLMs to assist in various stages of the design and verification process, improving efficiency and accuracy.

\textbf{Scope of This Work:} First, we explore LLMs for \emph{automated} testbench generation of FSMs.  
Through our case studies, we find that by \emph{using feedback from commercial EDA simulation tools}, LLMs can improve FSM test coverage. 
% and pin-point test cases that are crucial to check design specification violations.  
Second, we analyze the bug detection capability by \emph{providing output traces from the tools and utilizing prompting techniques}.
%By automating the testbench generation, LLMs can streamline the validation, reduce manual errors, and speed-up IC development.

% \begin{figure*}[tb]
% \centering
% \includegraphics[width=1.03\textwidth]{figures/ISLAD-Page-4.drawio.pdf}
% \caption{Framework for generation of LLM-assisted generation of testbenches with improved FSM coverage. (Correct the Figure,)
% }
% \label{fig:flow}

% \end{figure*}

\begin{figure}[tb]
\centering
\includegraphics[width=\columnwidth]{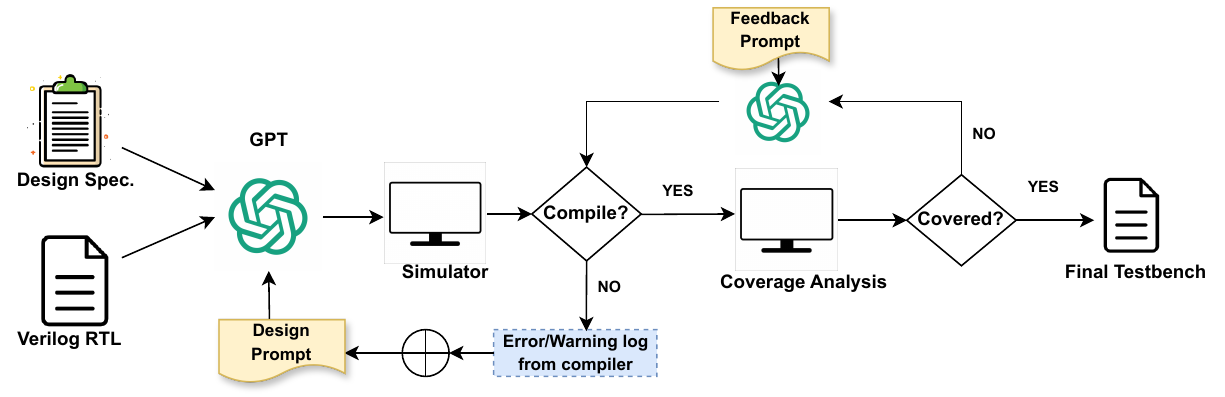}
\caption{\textit{Contribution 1:} LLM-aided Testbench Generation.}
\label{fig:coverage}

\end{figure}

\begin{figure}[tb]
\vspace*{-0.2in}
\centering
\includegraphics[width=\columnwidth]{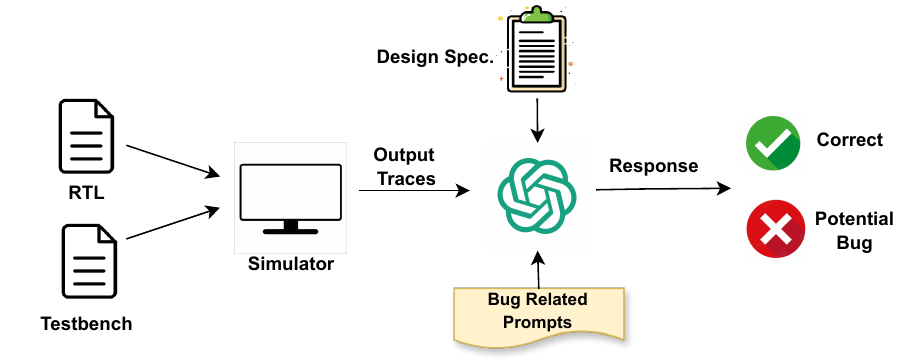}
\caption{\textit{Contribution 2:} LLM-aided Bug Detection.}
\label{fig:bugframe}
\vspace*{-0.2in}
\end{figure}

% \begin{figure*}[tb]
%     \centering
%     \subfloat[\label{fig:fixed_k}]{
%             \includegraphics[width=\columnwidth]{figures/ISLAD-Page-8.drawio.pdf}
%     }    
%     \subfloat[\label{fig:GateN4x4}]{
%         {\includegraphics[width=\columnwidth]{figures/ISLAD-Page-9.drawio.pdf}}
%     }%
%     \caption{}
% \end{figure*}
%\input{texfiles/2.Background}

\section{Background and Related Work\label{sec:background}}

Functional testing of RTL modules is crucial to ensure correct functionality. State-of-the-art verification techniques include both functional and formal verification, along with coverage analysis. These techniques rely on the careful formulation of assertions, often using System-Verilog, and the generation of comprehensive test patterns. Designers and test engineers find it especially challenging to create a testbench and generate effective test patterns. This difficulty increases when the design is still in progress and involves multiple team members. Testbench compilation requires a thorough understanding of the RTL code. Designers must meticulously analyze the code to generate test patterns that achieve complete coverage. Effective coverage analysis helps identify untested parts of the design, guiding further test development.

LLMs and their increasing prominence, as highlighted by the developments of models like GPT-4~\cite{openai2024gpt4}, have sparked broad interest in various fields. Specifically, chip design has seen a surge in innovative approaches leveraging LLMs, such as~\cite{zhong2023llm4eda}. Considerable progress has been made in improving the quality of Verilog code generation~\cite{liu2023verilogeval,thakur2023autochip,lu2023rtllm,thakur2023verigen}. These studies introduce methodologies to refine the process of generating Verilog code, demonstrating the potential of LLMs to streamline and improve hardware design workflows.
Furthermore, \cite{Blocklove_2023,fu2023gpt4aigchip,chang2023chipgpt} showcases the effectiveness of prompting strategies in chip design. They harness the power of LLMs to conceptualize and
detail the intricate aspects of hardware architectures, paving the way for efficient chip development.
The work in \cite{liu2023chipnemo} extends LLM applications into assistant chatbots, script generation, and bug analysis. Similarly, \cite{wu2024chateda} explores the use of LLMs for task planning and execution
within the EDA flow, highlighting the
potential for automating and optimizing complex workflows.
Finally, the generation of assertions for verifying the correctness of IC designs, has been enhanced through LLMs \cite{kande2023llmassisted,fang2024assertllm}.

% -------------------------------------------- %
%
% Fig 1. is currently in 2.Background.tex
% 
% -------------------------------------------- %

\section{Methodology \label{sec:framework}}

We study the capabilities of LLMs for both \emph{coverage-guided testbench generation} and \emph{automated bug detection}. 

\subsection{LLM (and coverage)-guided Testbench Generation}
\autoref{fig:coverage} illustrates an LLM-guided methodology for automatic testbench generation for FSMs. Any LLM can be employed and without loss of generality, we use GPT3.5 and GPT4. The inputs are: (1) an English-language specification\footnote{specification could be phrased as ``\emph{Write a Verilog module that detects a 1011 pattern and outputs a 1 anytime the last four inputs match with this pattern, and 0 otherwise.}''}, and
(2) a Verilog RTL description of design-under-test (DUT). 
% The DUT RTL represents the FSM that implements the given specification. Importantly, the RTL may contain bugs that need to be identified and addressed through the testing process.
The question we seek to address here is: \textit{Can we automatically generate a high-quality testbench for such DUT Verilog code?}
% \textcolor{red}{We should not include Step 1, i.e., the code description part in our paper. It's not implemented, evaluated or useful. I have commented it out.}
%
%To answer this question, 
We first need a metric to measure the quality of a testbench. A common metric, especially for FSMs, is \emph{transition coverage}, i.e., there should be at least one set of test patterns that covers all state transitions in the FSM. Several simulation tools provide \emph{coverage reports}, including state and transition coverage. Here, we investigate if we can use such coverage reports as feedback to iteratively improve coverage of an LLM-generated testbench.

Our method for automatic testbench generation involves several steps to ensure comprehensive testing of FSMs.
First, we prompt an LLM to generate an initial testbench using the Verilog code of the DUT as input, as shown in~\autoref{fig:prompt}(a). This Verilog code represents the FSM that needs to be tested. The LLM processes this input and generates a testbench that is intended to verify the function of the DUT.
Next, the generated testbench undergoes a compilation check using commercial EDA tools. In this step, the testbench is checked for syntax and logic errors. If any errors are detected during compilation, these errors are appended to the prompt and fed back to the LLM. The LLM uses this feedback to refine and correct its output iteratively until an error-free testbench is produced.
%Once an error-free testbench is generated,

Next, the FSM is simulated in coverage analysis mode. The simulation yields a report that quantifies how thoroughly the testbench exercises the FSM.
The coverage report identifies any FSM transitions that the testbench has missed, highlighting areas where the testbench needs improvement. Uncovered transitions in the coverage report are used as prompts for the LLM, as shown in~\autoref{fig:prompt}(b). These prompts guide the LLM to generate new test cases to cover missing transitions. By doing so, we ensure that the testbench comprehensively tests all functional parts of the FSM. Generating new test cases and performing coverage analysis is repeated, as shown in~\autoref{fig:coverage}. 
The cycle continues until full transition coverage is achieved %, meaning all possible transitions are tested. 
or may stop once a user-defined  threshold is met.
\autoref{fig:example} shows a testbench that yields 100\% coverage.

\begin{figure}[tb]
    \centering
\begin{subfigure}[b]{\columnwidth}

\begin{tcolorbox}[width=1.0\linewidth, halign=left, colframe=black, colback=white, boxsep=0.01mm, arc=1.5mm, left=2mm, right=2mm, boxrule=0.5pt]\footnotesize
You are an expert in design verification for Verilog code. Given a Verilog RTL module, you will write a testbench to simulate it and try to cover all the possible state transitions. Please follow the below instructions while providing any response: \\
\begin{enumerate}
    \item The testbench should start with \textit{module tb();} 
\item You will add \textit{\$fsdbDumpfile, \$fsdbDumpvars} commands in the testbench at the start of the first \textit{initial} block. 
\item Please use  \textit{apply\_input()} format to apply input sequences. 
\item You should consider whether it requires an active or high reset from the RTL code provided.
\item At the end of test patterns add \textit{\$finish}.
\end{enumerate}
 
\end{tcolorbox}

\vspace{-3mm}
    \label{fig:test_prompt1}
   \caption{System Prompt for GPT.}
\end{subfigure}

\begin{subfigure}[b]{\columnwidth}
\begin{tcolorbox}[width=1.0\linewidth, halign=left, colframe=black, colback=white, boxsep=0.01mm, arc=1.5mm, left=2mm, right=2mm, boxrule=0.5pt]\footnotesize
The above testbench provided doesn't cover all the transitions. This is the list of transitions that were expected but didn't happen:\\
\textit{"Transition from A to B"}\\
Please consider the  RTL Verilog code provided while providing the testbench and combine the test cases from the above response. You may have to reset a few times to cover certain transitions.
\end{tcolorbox}

\vspace{-3mm}
    \label{fig:test_prompt2}

   \caption{Prompt 1, to guide on some transitions not covered yet.}

\end{subfigure}

\begin{subfigure}[b]{\columnwidth}

\begin{tcolorbox}[width=1.0\linewidth, halign=left, colframe=black, colback=white, boxsep=0.01mm, arc=1.5mm, left=2mm, right=2mm, boxrule=0.5pt]\footnotesize
We ran the simulation tool with the testbench, this is the value for the state register variable across the clock cycle, the sequence is provided serially starting from 0 till the simulation finishes. This also shows the transition at each clock cycle.\\
\textit{"S0 S1 S2 S3 S1 S5...................................."}\\
 Please use the design specification provided and find out if there is any mismatch between them. We are looking to see if any transitions are inconsistent with the design spec.
\end{tcolorbox}
\vspace{-3mm}
    \label{fig:test_prompt3}
   \caption{Prompt 2, to verify the test outputs with the design specification.}

\end{subfigure}

\caption{Exemplary prompts used with GPT.}
\label{fig:prompt}

\vspace{-2mm}
\vspace*{-0.15in}

\end{figure}

\subsection{LLM-guided Automated Bug Detection}
RTL may contain bugs, either intentionally inserted or accidental, that need to be identified and fixed during testing. Given a testbench, we simulate the FSM and record the results of each test case. These simulation results and the original design specification in plain English are provided to the LLM. The LLM is tasked with identifying failing test cases by comparing the simulation outcomes with the expected behavior described in the specification in~\autoref{fig:bugframe}. This capability is powerful because the LLM can operate on less formal, natural-language descriptions. 
%This flexibility is advantageous since
Creating a formal specification is complex, and time-consuming.
~\autoref{fig:prompt}(c) provides prompts.
% demonstrating effectiveness of our method.

We use testbenches generated in part 1 as the starting point for bug detection. Our study shows that, as the FSM becomes complex the number of input patterns to detect the bugs increases notably. This leads to a large I/O trace obtained after the simulation, as shown in~\autoref{fig:bugframe}. GPT cannot comprehend this, thus missing
%not able to detect of
potential bugs. This naive approach thus does not scale.
To address these challenges, we employ
%improved results by employing
tww improvements to prompting: \textit{(1) Dividing I/O patterns into smaller sets to track states},
and \textit{(2) Handling multi-bit inputs and outputs individually to manage sub-circuits (output cones)},
%without overwhelming the LLM},
as shown in~\autoref{fig:prompt_4}. An analysis is in~\autoref{sec:exp}(C).

\begin{figure}[htbp]
    \centering
\begin{subfigure}[b]{\columnwidth}

\begin{tcolorbox}[width=1.0\linewidth, halign=left, colframe=black, colback=white, boxsep=0.01mm, arc=1.5mm, left=2mm, right=2mm, boxrule=0.5pt]\footnotesize
This is the input-output pair for the first 10 clock cycles. 
Please use the design specification provided and find out if there is any mismatch between them. We are looking to see if any transitions are inconsistent with the design spec. This will be followed up with the next 10 clock cycles and so on. After every 10th clock cycle please remember the current state which is very important when we provide new 10 input-output pairs.

\end{tcolorbox}

\vspace{-3mm}
    \label{fig:test_prompt5}
    %\vspace{-5mm}
   \caption{Prompt 3, sub-divide all the input/output pairs.}

\end{subfigure}

\begin{subfigure}[b]{0.98\columnwidth}

\begin{tcolorbox}[width=1.0\linewidth, halign=left, colframe=black, colback=white, boxsep=0.01mm, arc=1.5mm, left=2mm, right=2mm, boxrule=0.5pt]\footnotesize
To simplify the mismatch detection, consider the provided input-output pair for each clock cycle. Start the detection process by focusing on one bit of output and check for correct values as the input patterns are applied, followed by checking on other bits of output as well.

\end{tcolorbox}
\vspace{-3mm}
    \label{fig:test_prompt6}
    %\vspace{-5mm}
   \caption{Prompt 4, handle multi-bit input and output scenarios separately.}
\end{subfigure}
\caption{Additional Prompt for GPT during detection step. }
\label{fig:prompt_4}
\vspace{-2mm}
\end{figure}

%%%%%%%%%%%%%%%%%%%%%%%%%%%%%%%%%%%%%%%%%%%%%%%%%%%%%

\begin{figure}[htbp]
\centering
\begin{lstlisting}[language=verilog, basicstyle=\scriptsize,
% linebackgroundcolor={\ifnum\value{lstnumber}>5
%                 \ifnum\value{lstnumber}<9
%                     \color{pink}
%                 \fi
%             \fi}
            ]
module tb_fsm();
  reg clk;
  reg rst;
  reg inp;
  wire out1;
  wire out2;
  
  fsm uut(.clk(clk), .rst(rst), .inp(inp), .out1(out1), .out2(out2));
  // Clock generation
  always begin
    #5 clk = ~clk; 
  end

  // Test procedure
  initial begin
    // Initialize signals
    clk = 0;
    rst = 1;
    inp = 0;
    #10 rst = 0; 
    
    // Set up FSDB file for waveform viewing
    $fsdbDumpfile("test.fsdb");
    $fsdbDumpvars;

    // Test sequence to cover all transitions   
    // S0 -> S1 -> S2 -> S3 -> S4 -> S5 -> S6 -> S7 with inp=1
    apply_input_sequence(8'b11111111); 
   // Rest of the test patterns
   // .....
    $stop;
  end
  // Task to apply input and display results
  task apply_input(input reg in);
    inp = in;
    #10; // Wait for one clock cycle
  endtask
  // Task to apply a sequence of inputs
  task apply_input_sequence(input reg [7:0] seq);
    integer i;
    for (i = 7; i >= 0; i = i - 1) begin
      apply_input(seq[i]);
    end
  endtask

endmodule

\end{lstlisting}
\caption{An LLM-generated testbench exercising the design under test, in this instance an FSM.
%\textcolor{red}{this figure is too large, maybe keep only 2 input testcases?}
}
\label{fig:example}
\vspace{-2mm}
\end{figure}
\section{Experiments\label{sec:exp}}

\subsection{Setup}
Our study uses GPT(3.5/4), \textit{Synopsys VCS U-2023.03-1} and \textit{Synopsys Verdi U-2023.03-1}. The latter tools are for compiling and simulating Verilog code, debugging, and using its coverage reports as feedback. An example VCS report template is shown in \autoref{fig:report}. We automated the framework in \autoref{fig:coverage} and \autoref{fig:bugframe} in Python, including testbench generation, code compilation, coverage analysis, and comparison of test cases with natural-language specifications. For our case studies, we obtained a large number of representative FSMs from HDLBits~\cite{hdlbits} and GitHub, with varying complexity in terms of number of states and transitions. {100 FSMs}
are used in our study.\footnote{%
We show 50/100 cases, capturing the complexity of the dataset.
Full analysis of the datasets, including the RTL source code and the testbenches are available at~\cite{github}.
The remaining 50 cases are in the appendix, \autoref{tab:results_rest}.}
%For simplicity, we marked all FSMS serially. %numbers.

%%%%%%%%%%%%%%%%%%%%%%%%%%%%%%%%%%%%%%%%%

%%%%%%%%%%%%%%%%%%%%%%%%%%%%%%%%%%%%%%%%%%%%%%%%%%%%%%%%%%

\begin{figure}[htbp]
\center
\begin{tcolorbox}[width=0.96\linewidth, halign=left, colframe=black, colback=white, boxsep=0.01mm, arc=1.5mm, left=2mm, right=1mm, boxrule=0.5pt]\footnotesize

\begin{tabbing}

\texttt{FSM Coverage for Module : fsm} \\
\texttt{Summary for FSM :: current\_state} \\
\ \ \ \ \ \ \ \ \ \ \ \ \ \ \ \ \ \ \ \ \ \textbf{Total} \ \ \ \ \ \ \textbf{Covered} \ \ \ \ \ \textbf{Percent} \\
States \ \ \ \ \ \ \ \ \ \ \ \ \ \ \ \    4 \ \ \ \ \ \ \ \ \ \ \  4 \ \ \ \ \ \ \ \ \ \ 100.00 \\
\textcolor{red}{Transitions} \ \ \ \ \ \ \ \ \ \ \ 8 \ \ \ \ \ \ \ \ \ \ \ 6 \ \ \ \ \ \ \ \ \ \ \ \textcolor{red}{75.00} \\
% Sequences \ \ \ \ \ \ \ \ \ \ \ 0 \ \ \ \ \ \ \ 0 \ \ \ \ \ \ \ \ 0.00 \\
% State, Transition and Sequence Details for FSM :: current\_state \\
% ---------------------------------------------------------------- 
\textbf{States} \ \ \ \ \ \ \ \ \ \textbf{Line No.} \ \ \ \ \  \textbf{Covered} \\
A \ \ \ \ \ \ \ \ \ \ \ \ \ \ \ \ 17 \ \ \ \ \ \ \ \ \ \  Covered \\
B \ \ \ \ \ \ \ \ \ \ \ \ \ \ \ \ 29 \ \ \ \ \ \ \ \ \ \ Covered \\
C \ \ \ \ \ \ \ \ \ \ \ \ \ \ \ \ 38 \ \ \ \ \ \ \ \ \ \ Covered \\
D \ \ \ \ \ \ \ \ \ \ \ \ \ \ \ \ 35 \ \ \ \ \ \ \ \ \ \ Covered \\
\textbf{Transitions} \ \ \ \ \ \  \textbf{Line No.} \ \ \ \ \ \ \ \ \ \  \textbf{Covered} \\
A$\rightarrow$A \ \ \ \ \ \ \ \ \ \ \ \ \ \ 17 \ \ \ \ \ \ \ \ \ \ \ \ \ \ \ Covered \\
A$\rightarrow$B \ \ \ \ \ \ \ \ \ \ \ \ \ \ 29 \ \ \ \ \ \ \ \ \ \ \ \ \ \ \ Covered \\
\textcolor{red}{B$\rightarrow$A} \ \ \ \ \ \ \ \ \ \ \ \ \ \ \textcolor{red}{17} \ \ \ \ \ \ \ \ \ \ \ \ \ \ \ \textcolor{red}{Not Covered} \\
B$\rightarrow$C \ \ \ \ \ \ \ \ \ \ \ \ \ \ 38 \ \ \ \ \ \ \ \ \ \ \ \ \ \ \ Covered \\
B$\rightarrow$D \ \ \ \ \ \ \ \ \ \ \ \ \ \ 35 \ \ \ \ \ \ \ \ \ \ \ \ \ \ \ Covered \\
\textcolor{red}{C$\rightarrow$A} \ \ \ \ \ \ \ \ \ \ \ \ \ \ \textcolor{red}{17} \ \ \ \ \ \ \ \ \ \ \ \ \ \ \ \textcolor{red}{Not Covered} \\
C$\rightarrow$D \ \ \ \ \ \ \ \ \ \ \ \ \ \ 44 \ \ \ \ \ \ \ \ \ \ \ \ \ \ \ Covered \\
D$\rightarrow$A \ \ \ \ \ \ \ \ \ \ \ \ \ \ 17 \ \ \ \ \ \ \ \ \ \ \ \ \ \ \ Covered 
\end{tabbing}

\end{tcolorbox}

\caption{FSM coverage report template.}
\label{fig:report}
\end{figure}
%%%%%%%%%%%%%%%%%%%%%%%%%%%%%%%%%%%%%%%%%%%%%%%%%%%%%%%%%%
\begin{figure}[!t]
%\vspace*{-0.2in}
\centering
\subfloat[\label{fig:original} Correct]{\includegraphics[width=0.52\columnwidth]{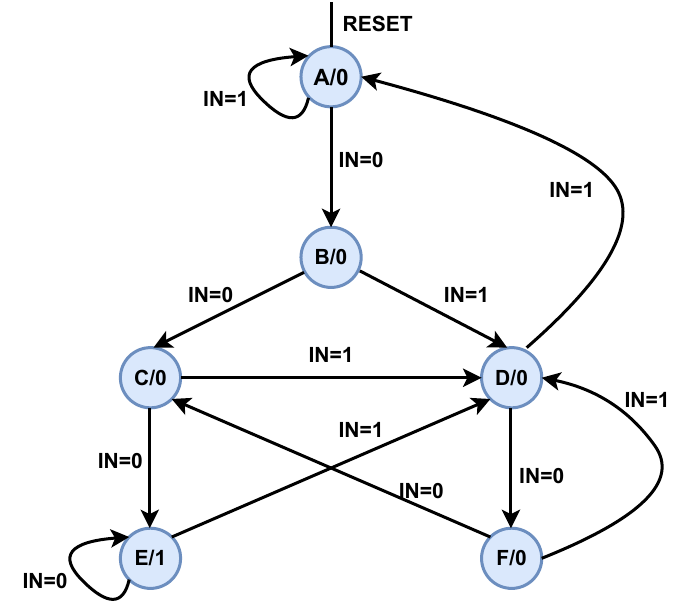}} 
\subfloat[\label{fig:bug} Buggy]
{\includegraphics[width=0.52\columnwidth]{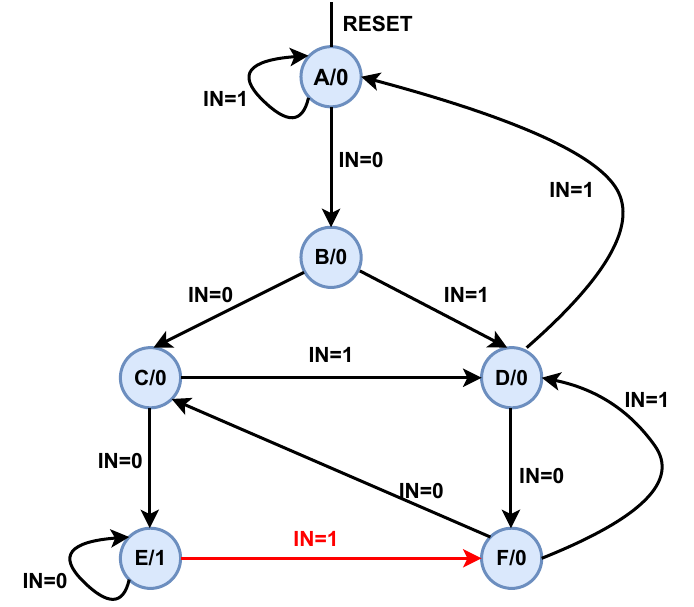}} 
\caption{State transition diagram for FSM16. }
\label{fig:fig1}
\vspace*{-0.2in}
\end{figure}

\begin{table*}[!t]

\caption{Results on FSMs of different complexity. All FSMs have natural-language prompts describing the function. "Iters" are the number of iterations required to achieve \% Cov (Coverage). "State Regs", "I/O pairs" and "Fuzzing" denote \# input patterns required to detect a mismatch from the specification by the three methods, respectively. FSMs are arranged from 1--100. 50 FSMs are reported here, rest are reported in Appendix~\ref{app:case}.  
(\checkmark) denotes successful bug detection  and (\xmark) denotes failed attempt.
}
\label{tab:results_our}
\centering
\setlength{\tabcolsep}{2.5pt}
\resizebox{\textwidth}{!}{
\begin{tabular}{|c|c|c|c|c|c|c|c|c|c|c|c|c|c|c|c|c|c|c|c|c|c|c|}
\hline
\multirow{7}{*}{ \rotatebox{90}{\textbf{Level}} } & \multirow{7}{*}{ \rotatebox{90}{\textbf{FSMs}} } & \multicolumn{3}{c|}{\multirow{3}{*}{ \textbf{Characteristics} }} & \multicolumn{6}{c|}{\textbf{LLM (and coverage)-guided Testbench Generation}} & \multicolumn{12}{c|}{\textbf{LLM-guided Automated Bug Detection}} \\ \cline{6-23}
& & \multicolumn{3}{c|}{} & 
{ \textbf{GPT3.5}} & 
{ \textbf{GPT4}} & \multicolumn{2}{c|}{\textbf{GPT3.5+This Work}} & \multicolumn{2}{c|}{\textbf{GPT4+This Work}} & \multicolumn{3}{c|}{ \textbf{GPT3.5} } & \multicolumn{3}{c|}
{ \textbf{GPT4} } & \multicolumn{3}{c|}{ \textbf{GPT3.5+This Work}} & \multicolumn{3}{c|}{\textbf{GPT4+This Work }} \\ 
\cline{3-23}
& & \multirow{4}{*}{ \rotatebox{90}{i/o} } & \multirow{4}{*}{ \rotatebox{90}{o/p} } & \multirow{4}{*}{ \rotatebox{90}{states} } & \multirow{4}{*}{ \rotatebox{90}{\% Cov} } & \multirow{4}{*}{ \rotatebox{90}{\% Cov} } & \multirow{4}{*}{ \rotatebox{90}{Iters} } & \multirow{4}{*}{ \rotatebox{90}{\% Cov} } & \multirow{4}{*}{ \rotatebox{90}{Iters} } & \multirow{4}{*}{ \rotatebox{90}{\% Cov}}  & \multirow{4}{*}{ \rotatebox{90}{State Regs} } & \multirow{4}{*}{ \rotatebox{90}{I/O pairs} } & \multirow{4}{*}{ \rotatebox{90}{Fuzzing} } & \multirow{4}{*}{ \rotatebox{90}{State Regs} } & \multirow{4}{*}{ \rotatebox{90}{I/O pairs} } & \multirow{4}{*}{ \rotatebox{90}{Fuzzing} } & \multirow{4}{*}{ \rotatebox{90}{State Regs} } & \multirow{4}{*}{ \rotatebox{90}{I/O pairs} } & \multirow{4}{*}{ \rotatebox{90}{Fuzzing} } & \multirow{4}{*}{ \rotatebox{90}{State Regs} } & \multirow{4}{*}{ \rotatebox{90}{I/O pairs} } & \multirow{4}{*}{ \rotatebox{90}{Fuzzing} } \\
& & & & & & & & & & & & & & & & & & & & & & \\
& & & & & & & & & & & & & & & & & & & & & & \\
& & & & & & & & & & & & & & & & & & & & & & \\ \hline
\multirow{24}{*}{ \rotatebox{90}{Easy} } 
& 1 & 2 & 1 & 2 & 50 & 50 & 2 & 100 & 2 & 100 & \checkmark & \checkmark & \checkmark & \checkmark & \checkmark & \checkmark & \checkmark & \checkmark & \checkmark & \checkmark & \checkmark & \checkmark \\
& 2 & 6 & 3 & 2 & 75 & 50 & 2 & 100 & 2 & 100 & \checkmark & \checkmark & \checkmark & \checkmark & \checkmark & \checkmark & \checkmark & \checkmark & \checkmark & \checkmark & \checkmark & \checkmark \\
& 9 & 1 & 1 & 3 & 33 & 75 & 3 & 100 & 2 & 100 & \checkmark & \checkmark & \checkmark & \checkmark & \checkmark & \checkmark & \checkmark & \checkmark & \checkmark & \checkmark & \checkmark & \checkmark \\
& 10 & 1 & 1 & 3 & 50 & 50 & 3 & 100 & 2 & 100 & \checkmark & \checkmark & \checkmark & \checkmark & \checkmark & \checkmark & \checkmark & \checkmark & \checkmark & \checkmark & \checkmark & \checkmark \\
& 15 & 3 & 3 & 4 & 75 & 50 & 3 & 100 & 3 & 100 & \checkmark & \checkmark & \checkmark & \checkmark & \checkmark & \checkmark & \checkmark & \checkmark & \checkmark & \checkmark & \checkmark & \checkmark \\
& 16 & 2 & 2 & 4 & 25 & 50 & 3 & 100 & 2 & 100 & \checkmark & \checkmark & \checkmark & \checkmark & \checkmark & \checkmark & \checkmark & \checkmark & \checkmark & \checkmark & \checkmark & \checkmark \\
& 17 & 2 & 1 & 4 & 25 & 50 & 2 & 90 & 2 & 100 & \checkmark & \checkmark & \checkmark & \checkmark & \checkmark & \checkmark & \checkmark & \checkmark & \checkmark & \checkmark & \checkmark & \checkmark \\
& 18 & 2 & 8 & 4 & 25 & 50 & 4 & 100 & 2 & 100 & \checkmark & \checkmark & \checkmark & \checkmark & \checkmark & \checkmark & \checkmark & \checkmark & \checkmark & \checkmark & \checkmark & \checkmark \\
& 19 & 5 & 2 & 4 & 25 & 50 & 5 & 90 & 2 & 100 & \checkmark & \checkmark & \checkmark & \checkmark & \checkmark & \checkmark & \checkmark & \checkmark & \checkmark & \checkmark & \checkmark & \checkmark \\
& 20 & 3 & 3 & 4 & 50 & 50 & 3 & 100 & 2 & 100 & \checkmark & \checkmark & \checkmark & \checkmark & \checkmark & \checkmark & \checkmark & \checkmark & \checkmark & \checkmark & \checkmark & \checkmark \\
& 25 & 1 & 1 & 5 & 25 & 20 & 5 & 90 & 3 & 100 & \checkmark & \checkmark & \checkmark & \checkmark & \checkmark & \checkmark & \checkmark & \checkmark & \checkmark & \checkmark & \checkmark & \checkmark \\
& 26 & 2 & 8 & 5 & 33 & 50 & 6 & 100 & 2 & 100 & \checkmark & \checkmark & \checkmark & \checkmark & \checkmark & \checkmark & \checkmark & \checkmark & \checkmark & \checkmark & \checkmark & \checkmark \\
& 27 & 1 & 2 & 5 & 25 & 35 & 5 & 100 & 3 & 90 & \checkmark & \checkmark & \checkmark & \checkmark & \checkmark & \checkmark & \checkmark & \checkmark & \checkmark & \checkmark & \checkmark & \checkmark \\
& 28 & 3 & 3 & 5 & 50 & 50 & 4 & 95 & 3 & 100 & \checkmark & \checkmark & \checkmark & \checkmark & \checkmark & \checkmark & \checkmark & \checkmark & \checkmark & \checkmark & \checkmark & \checkmark \\
& 29 & 2 & 6 & 5 & 50 & 50 & 4 & 90 & 3 & 100 & \checkmark & \checkmark & \checkmark & \checkmark & \checkmark & \checkmark & \checkmark & \checkmark & \checkmark & \checkmark & \checkmark & \checkmark \\
& 30 & 1 & 1 & 6 & 25 & 25 & 5 & 90 & 2 & 100 & \checkmark & \checkmark & \checkmark & \checkmark & \checkmark & \checkmark & \checkmark & \checkmark & \checkmark & \checkmark & \checkmark & \checkmark \\
& 31 & 6 & 2 & 6 & 25 & 20 & 6 & 100 & 2 & 100 & \checkmark & \checkmark & \checkmark & \checkmark & \checkmark & \checkmark & \checkmark & \checkmark & \checkmark & \checkmark & \checkmark & \checkmark \\
& 32 & 7 & 3 & 6 & 20 & 25 & 4 & 100 & 4 & 100 & \checkmark & \checkmark & \checkmark & \checkmark & \checkmark & \checkmark & \checkmark & \checkmark & \checkmark & \checkmark & \checkmark & \checkmark \\
& 33 & 4 & 4 & 6 & 43 & 37 & 6 & 90 & 3 & 100 & \checkmark & \checkmark & \checkmark & \checkmark & \checkmark & \checkmark & \checkmark & \checkmark & \checkmark & \checkmark & \checkmark & \checkmark \\
& 34 & 1 & 1 & 6 & 23 & 32 & 6 & 95 & 3 & 95 & \checkmark & \checkmark & \checkmark & \checkmark & \checkmark & \checkmark & \checkmark & \checkmark & \checkmark & \checkmark & \checkmark & \checkmark \\
& 35 & 2 & 4 & 6 & 40 & 35 & 5 & 100 & 3 & 90 & \checkmark & \checkmark & \checkmark & \checkmark & \checkmark & \checkmark & \checkmark & \checkmark & \checkmark & \checkmark & \checkmark & \checkmark \\
& 36 & 1 & 2 & 6 & 21 & 40 & 5 & 90 & 3 & 100 & \checkmark & \checkmark & \checkmark & \checkmark & \checkmark & \checkmark & \checkmark & \checkmark & \checkmark & \checkmark & \checkmark & \checkmark \\
& 42 & 5 & 5 & 7 & 30 & 19 & 7 & 90 & 5 & 100 & \checkmark & \checkmark & \checkmark & \checkmark & \checkmark & \checkmark & \checkmark & \checkmark & \checkmark & \checkmark & \checkmark & \checkmark \\
& 43 & 4 & 4 & 7 & 20 & 50 & 6 & 92 & 3 & 100 & \checkmark & \checkmark & \checkmark & \checkmark & \checkmark & \checkmark & \checkmark & \checkmark & \checkmark & \checkmark & \checkmark & \checkmark \\ \hline
\multirow{16}{*}{ \rotatebox{90}{Medium} } 
& 46 & 3 & 7 & 8 & 50 & 33 & 8 & 96 & 6 & 95 & \checkmark & \checkmark & \checkmark & \checkmark & \checkmark & \checkmark & \checkmark & \checkmark & \checkmark & \checkmark & \checkmark & \checkmark \\
& 47 & 3 & 7 & 8 & 12 & 24 & 7 & 90 & 5 & 90 & \checkmark & \checkmark & \checkmark & \checkmark & \checkmark & \checkmark & \checkmark & \checkmark & \checkmark & \checkmark & \checkmark & \checkmark \\
& 48 & 1 & 2 & 8 & 20 & 13 & 7 & 91 & 7 & 95 & \checkmark & \checkmark & \checkmark & \checkmark & \checkmark & \checkmark & \checkmark & \checkmark & \checkmark & \checkmark & \checkmark & \checkmark \\
& 52 & 2 & 2 & 9 & 25 & 15 & 8 & 90 & 5 & 95 & \checkmark & \checkmark & \checkmark & \checkmark & \checkmark & \checkmark & \checkmark & \checkmark & \checkmark & \checkmark & \checkmark & \checkmark \\
& 53 & 2 & 4 & 9 & 10 & 12 & 9 & 92 & 8 & 100 & \checkmark & \xmark & \checkmark & \checkmark & \checkmark & \checkmark & \checkmark & \checkmark & \checkmark & \checkmark & \checkmark & \checkmark \\
& 54 & 1 & 1 & 9 & 19 & 33 & 7 & 95 & 6 & 92 & \checkmark & \xmark & \checkmark & \checkmark & \checkmark & \checkmark & \checkmark & \checkmark & \checkmark & \checkmark & \checkmark & \checkmark \\
& 59 & 2 & 2 & 10 & 8 & 10 & 9 & 95 & 5 & 90 & \checkmark & \xmark & \xmark & \checkmark & \checkmark & \checkmark & \checkmark & \checkmark & \checkmark & \checkmark & \checkmark & \checkmark \\
& 60 & 1 & 3 & 10 & 11 & 15 & 8 & 100 & 8 & 100 & \checkmark & \xmark & \xmark & \checkmark & \checkmark & \checkmark & \checkmark & \checkmark & \checkmark & \checkmark & \checkmark & \checkmark \\
& 61 & 3 & 3 & 10 & 25 & 20 & 8 & 90 & 8 & 90 & \checkmark & \xmark & \xmark & \checkmark & \checkmark & \checkmark & \checkmark & \checkmark & \checkmark & \checkmark & \checkmark & \checkmark \\
& 62 & 2 & 1 & 10 & 15 & 12 & 8 & 100 & 8 & 100 & \xmark & \xmark & \xmark & \checkmark & \checkmark & \checkmark & \checkmark & \checkmark & \checkmark & \checkmark & \checkmark & \checkmark \\
& 66 & 6 & 5 & 11 & 13 & 10 & 10 & 90 & 9 & 90 & \xmark & \xmark & \xmark & \checkmark & \checkmark & \checkmark & \checkmark & \checkmark & \checkmark & \checkmark & \checkmark & \checkmark \\
& 67 & 1 & 1 & 11 & 7 & 11 & 9 & 92 & 9 & 95 & \xmark & \xmark & \xmark & \checkmark & \xmark & \checkmark & \checkmark & \checkmark & \checkmark & \checkmark & \checkmark & \checkmark \\
& 71 & 12 & 3 & 12 & 22 & 13 & 12 & 90 & 8 & 90 & \xmark & \xmark & \xmark & \checkmark & \xmark & \xmark & \checkmark & \checkmark & \checkmark & \checkmark & \checkmark & \checkmark \\
& 72 & 2 & 1 & 12 & 13 & 12 & 9 & 94 & 9 & 95 & \xmark & \xmark & \xmark & \checkmark & \xmark & \xmark & \checkmark & \checkmark & \checkmark & \checkmark & \checkmark & \checkmark \\
& 75 & 1 & 11 & 13 & 18 & 10 & 13 & 92 & 9 & 90 & \xmark & \xmark & \xmark & \xmark & \xmark & \xmark & \checkmark & \checkmark & \checkmark & \checkmark & \checkmark & \checkmark \\
& 78 & 4 & 16 & 14 & 6 & 7 & 14 & 90 & 10 & 100 & \xmark & \xmark & \xmark & \xmark & \xmark & \xmark & \checkmark & \checkmark & \checkmark & \checkmark & \checkmark & \checkmark \\
\hline
\multirow{10}{*}{ \rotatebox{90}{Hard} } 
& 82 & 1 & 6 & 16 & 9 & 12 & 15 & 95 & 11 & 95 & \xmark & \xmark & \xmark & \xmark & \xmark & \xmark & \checkmark & \xmark & \checkmark & \checkmark & \checkmark & \checkmark \\
& 85 & 1 & 1 & 19 & 7 & 5 & 18 & 91 & 13 & 93 & \xmark & \xmark & \xmark & \xmark & \xmark & \xmark & \checkmark & \xmark & \xmark & \checkmark & \checkmark & \checkmark \\
& 87 & 2 & 2 & 20 & 5 & 6 & 20 & 90 & 13 & 90 & \xmark & \xmark & \xmark & \xmark & \xmark & \xmark & \xmark & \xmark & \xmark & \checkmark & \checkmark & \checkmark \\
& 90 & 1 & 1 & 22 & 7 & 7 & 21 & 90 & 14 & 94 & \xmark & \xmark & \xmark & \xmark & \xmark & \xmark & \xmark & \xmark & \xmark & \checkmark & \checkmark & \checkmark \\
& 92 & 4 & 4 & 23 & 8 & 11 & 22 & 91 & 12 & 93 & \xmark & \xmark & \xmark & \xmark & \xmark & \xmark & \xmark & \xmark & \xmark & \checkmark & \checkmark & \checkmark \\
& 93 & 1 & 2 & 24 & 4 & 10 & 23 & 94 & 13 & 91 & \xmark & \xmark & \xmark & \xmark & \xmark & \xmark & \xmark & \xmark & \xmark & \checkmark & \checkmark & \checkmark \\
& 95 & 2 & 1 & 25 & 8 & 12 & 22 & 90 & 11 & 92 & \xmark & \xmark & \xmark & \xmark & \xmark & \xmark & \xmark & \xmark & \xmark & \checkmark & \checkmark & \checkmark \\
& 96 & 1 & 1 & 25 & 5 & 15 & 23 & 93 & 12 & 100 & \xmark & \xmark & \xmark & \xmark & \xmark & \xmark & \xmark & \xmark & \xmark & \checkmark & \checkmark & \checkmark \\
& 99 & 2 & 2 & 27 & 12 & 9 & 24 & 95 & 13 & 92 & \xmark & \xmark & \xmark & \xmark & \xmark & \xmark & \xmark & \xmark & \xmark & \checkmark & \checkmark & \checkmark \\
& 100 & 6 & 16 & 28 & 6 & 5 & 22 & 95 & 14 & 95 & \xmark & \xmark & \xmark & \xmark & \xmark & \xmark & \xmark & \xmark & \xmark & \checkmark & \checkmark & \checkmark \\
\hline
\end{tabular}
}
\vspace*{-0.2in}
\end{table*}
%%%%%%%%%%%%%%%%%%%%%%%%%%%%%%%%%%%%%%%%%%%%%%%%%%%%

\subsection{Results: LLM-guided Testbench Generation}
To understand the limitations of using LLMs, if any, we have first studied the obtained testbenches in detail, providing the following important insights.
First, for certain paths and transitions selected for extended coverage, the model may not be able to figure out the next correct input pattern to move to the next state.
%\footnote{%
%\autoref{fig:original} shows a state transition diagram for  FSM16.
For example, to cover all  transitions in \autoref{fig:original} (state transition diagram for FSM16), i.e., A$\rightarrow$A, A$\rightarrow$B, B$\rightarrow$C, B$\rightarrow$D, C$\rightarrow$D, C$\rightarrow$E, E$\rightarrow$D, E$\rightarrow$E,
     D$\rightarrow$F, F$\rightarrow$C, F$\rightarrow$D, and D$\rightarrow$A, we will require an input pattern that begins with coverage of state A, like 1000010010101101....
This requirement was not automatically understood/identified by GPTs, so the model could get stuck at some state and repeat generating the same test patterns.
     %, without using some reset signal again.
Second, due to the context limits of LLM for that matter, the model cannot cover all transitions in one iteration. Thus, we prompt it to `reset' after certain transitions such that it can discover all input patterns up to some state transitions and traverse the rest, as shown in~\autoref{fig:prompt}.

\autoref{tab:results_our} provides representative results for our method in~\autoref{sec:framework}. As indicated, further results are made available at~\cite{github}.
%\footnote{%
%Due to page limits, we only cover a selected subset of all considered FSM designs in 
%\autoref{tab:results_our}.
%}
Our dataset has \textit{\{Easy, Medium, Difficult\}} FSMs, based on their complexity in terms of the number of states and the number of transitions (which is exponentially related to the number of states). The dataset showcases the capabilities of LLMs  and their limitations (e.g., due to a lack of domain knowledge).
First, we characterized the FSMs based on the number of \textit{inputs (i/p)}, \textit{outputs (o/p)}, and \textit{states}. \textit{Iter}ations relate to our method explained in~\autoref{sec:framework}; each iteration involves running VCS and collecting key information, as highlighted in~\autoref{fig:report}, which includes the percentage of coverage and 'Not Covered' transitions with the current test cases. This information is used to provide more context for the next prompt.
Second, we report coverage with no feedback, where we prompted \textit{\{GPT3.5, GPT4\}} to generate a testbench with full coverage. Correspondingly, \textit{\{GPT3.5 + This Work, GPT4 + This Work\}} represents coverage using feedback with our method. Varying the number of iterations is required for testbench generation across  FSMs. This is expected as the iterative process depends on the number of states, transitions, and the type of connections. Due to repetitive responses after some iterations and certain transitions being very difficult for the LLMs to understand, our aim was to reach as close to 100\% coverage as possible. As shown in~\autoref{tab:results_our}, with our feedback mechanism we were able to achieve that with both GPT3.5 and GPT4. Still, the number of iterations required to achieve that is less for GPT4 versus GPT3.5,
%varies across GPT3.5 and GPT4, with GPT4 having lesser number compared to GPT3.5,
which is expected considering the capabilities of both models.
For example, for FSM16, without feedback, the coverage is 25\% for both GPT3.5 and GPT4, whereas for our method it is 100\% in both cases.

%%%%%%%%%%%%%%%%%%%%%%%%%%%%%%%%%%%%%%%%%%%%%%%%%%%%%%

%%%%%%%%%%%%%%%%%%%%%%%%%%%%%%%%%%%%%%%%%%%%%%%%%%%%%%%
\subsection{Results: LLM-guided Automatic Bug Detection}
\label{case_study}

So far, we did \underline{\textit{not}} provide natural-language design specifications for any of the FSMs. Next we investigate automatic comparison of the natural-language specification against the generated test cases and simulation outputs. Such capabilities are essential for bug detection. Consider the `buggy' and incorrect transition in FSM16 in~\autoref{fig:bug} that occured due to human designer error.  

As shown in~\autoref{fig:bugframe},  we provide the tool with the test patterns and the VCS simulation outputs for each clock cycle. We prompt GPT(3.5/4) to correspond/match these test results with the design specifications and to find mismatches, if any. The mismatches are highlighted as potential bugs in the FSM RTL. 
%%replacing with text below since it is similar
We compare three scenarios in~\autoref{tab:results_our}:
(1) \textit{State Regs (Registers):} We provide all input patterns and state register values for each clock cycle; (2) \textit{I/O pairs:} We provide only iI?O pairs per clock cycle; (3) \textit{Fuzzing:} We provide random input patterns. For the Scenarios (1, 2), we applied our framework to generate test patterns. Scenario (3) is a baseline non-AI approach akin to hardware \emph{fuzzing}. 

\textbf{\textit{Initial Findings:}} LLMs can help raise warnings for mismatches in an advanced manner beyond traditional coverage.\footnote{%
100\% transition coverage does not guarantee bug detection.
%, especially for Scenario (2).
Consider `buggy' transition in FSM16 in~\autoref{fig:bug}. A test case reaches the incorrect transition but goes no further and fails to find the bug. FSM output diverges
%from its correct behavior
only a few cycles later.}
However, when using large testbenches, it can become difficult for the model to follow up for longer sequences of input patterns. Looking at each case in more detail for \textit{\{GPT3.5, GPT4\}} in the~\autoref{tab:results_our}, we find that in
Scenario (1), providing the state registers value helps the LLM to correlate with the design specification. That is, the model only needs to check for the next state based on the input patterns, and any transitions that are not mentioned/covered would raise a warning of a potential bug. For this scenario, thus, we can detect incorrect transitions for a few of the \textit{Easy} and \textit{Medium} level FSMs with a reasonable number of input patterns, but not for the \textit{Hard} FSMs. This is  because of the inability of these models to comprehend a large input context, with GPT4 performing better compared to GPT3.5.
Scenario (2) is challenging. We had to add some additional input patterns, thus leading to a bigger output trace to be fed to these models. We observe a similar trend to Scenario (1) across GPT3.5 and GPT4.
Scenario (3), i.e., random test pattern generation, performs similarly to the other 2 scenarios, because of the mentioned limitation of these models.
% GPT-4 could not comprehend the input sequences in their entirety, which is due to limited context length of LLMs.

\begin{figure*}[htbp]
\vspace*{-0.3in}
\centering
\includegraphics[width=\textwidth]{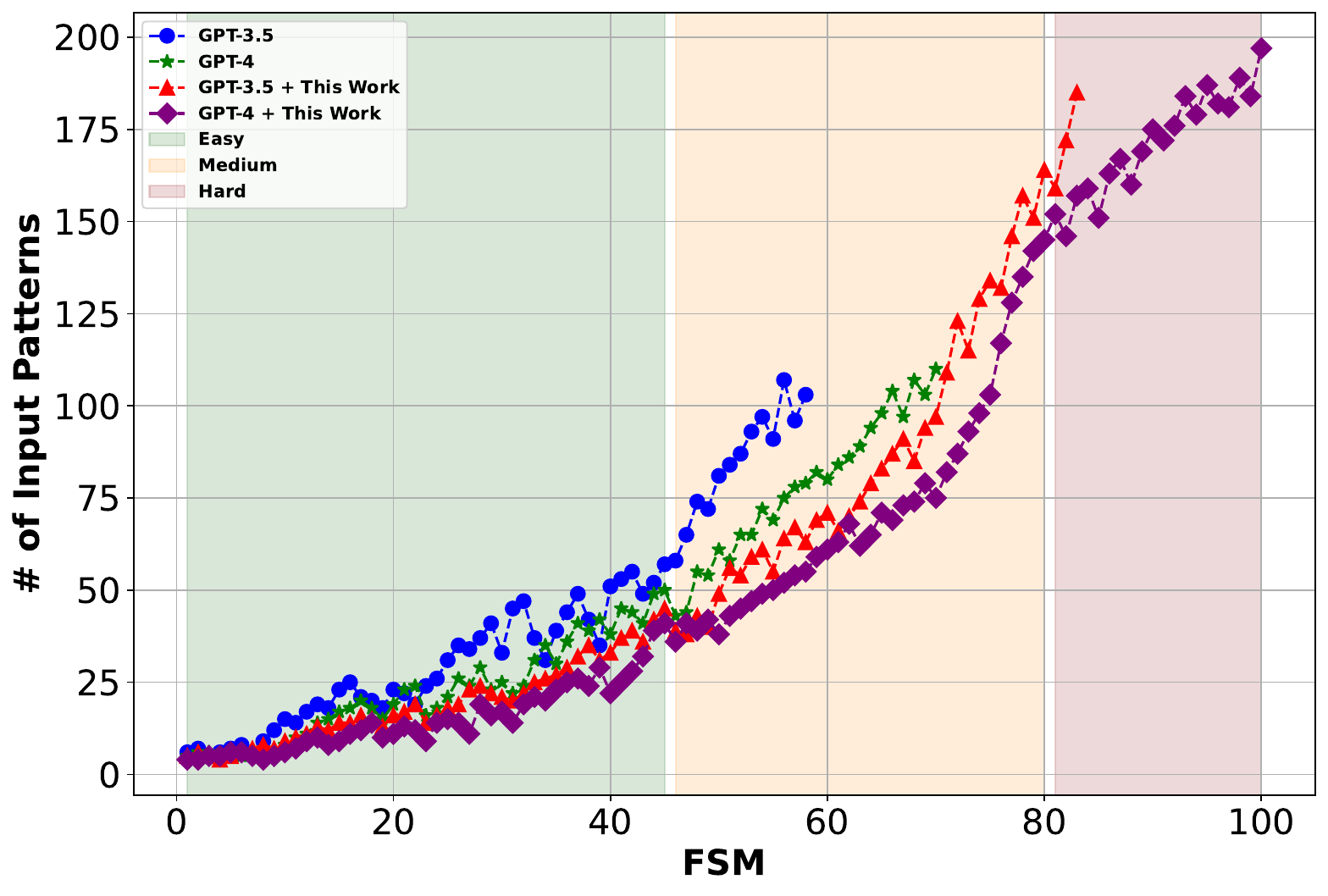}
%\vspace*{-0.2in}
\caption{\# of input patterns for bug detection for 100 FSMs using \textit{fuzzing} for all methods from~\autoref{tab:results_our}. Results are arranged as regions of FSM complexity. For missing points, respective methods could not detect the bug.}
\label{fig:plot}
\vspace*{-0.2in}
\end{figure*}

\textbf{\textit{Analysis for Bug Detection:}} We explored other approaches to improve the result, by adjusting the prompt as shown in \autoref{fig:prompt_4}. In one technique, instead of feeding all input-output patterns, we divide them into smaller sets. This is easier to handle but still allows the LLM to track the state reached at the end of each set. The next iteration of our method proceeds from there.
Another technique applies to multi-bit inputs and outputs. Instead of considering all output bits at once, we tackle them individually. That is, we provided all input patterns for an output bit and repeated this for all outputs. The rationale is to focus LLMs on manageable sub-circuits (output cones) rather than ``bombarding'' them with all information. We provide the natural-language description with each query to LLM, such that the model does not lose the overall context.
In \autoref{tab:results_our}, the columns corresponding to \textit{\{GPT3.5 + This Work, GPT4 + This Work\}} reports our final results with all these prompting techniques. With GPT3.5 and our method, we were able to detect the bug in all of \textit{Easy} and \textit{Medium} FSMs across all the 3 scenarios and a few \textit{Hard} FSMs for Scenarios (1) and (3). GPT4 combined with our method can detect the bug in all the 3 level of FSMs and for all scenarios. This shows the effectiveness of our method to improve the performance of LLMs on this task.
\autoref{fig:plot} shows the scaling of input patterns required to detect bugs for the \textit{fuzzing} approach for all methods. % in this study.
%we have observed a similar trend for the other 2 scenarios.
Clearly, our method combined with GPT3.5 and GPT4 can detect bugs with fewer patterns and with a higher success rate. % across the 100 FSMs.

% \begin{figure}[tb]
% \centering
% \includegraphics[width=\columnwidth]{figures/chart.pdf}
% \caption{Number of Input patterns required for detection of the bug for whole datasets across different scenario for \textit{\{GPT4 + This Work\}}, the best performing case from the~\autoref{tab:results_our}.}
% \label{fig:plot}

% \end{figure}

\subsection{Detailed Case Studies using GPT4}
%Here, we analyze few of the FSMs with GPT4 in more detail.

\subsubsection{FSM22}
This FSM has one input, multiple outputs, and 6 states. For Scenarios (1) and (2), with less number of patterns, it was easy for the model to analyze and flag the mismatch between the generated output and the design specification.
For Scenario (3), we employ new prompts similar to those shown in ~\autoref{fig:prompt_4}. The mismatch was detected with 64 patterns. 
\subsubsection{FSM34}
has 2 inputs, 1 output, and 10 states.
For Scenarios (1,2), we detected incorrect transitions by accessing state register and I/O values, respectively, for each clock cycle. For Scenario (3) using random patterns, GPT4 could not comprehend the input sequences.  
We modified the prompt similar to~\autoref{fig:prompt_4}, i.e.,
we iteratively consider subsets of I/O pairs. We begin with 100 random patterns and run the simulator to collect related I/O pairs, which are then fed to GPT4 in small chunks. GPT4 then detected a mismatch with $\sim$89 patterns. 
\subsubsection{FSM42}
 This is a 1-bit input, 1-bit output, and 19 state FSM. %Feeding all the input/output pairs to GPT-4 did not achieve bug detection. For all the cases, GPT-4 could not comprehend the whole FSM.
 For Scenarios (1) and (2), 73 and 85 patterns, respectively were needed to detect the mismatch. For Scenario (3), feeding all I/O pairs to GPT4 did not catch the bug.
 Thus, we used the prompts from ~\autoref{fig:prompt_4} and increased the random patterns to 200. A mismatch was detected  with $\sim$107 patterns. 

\subsubsection{FSM50}
is complex, with multi-bit outputs and 28 states. Our approach of feeding all the input-output pairs failed for all the cases. Instead, we collected 200 patterns for each case, where GPT4 still failed to comprehend the whole FSM. We employed both prompt variations in~\autoref{fig:prompt_4}.
For Scenario (1), with $\sim$158 test patterns and full access to state register values.
For Scenarios (2,3), for this revised approach,  GPT-4 detected the mismatch with $\sim$193 and $\sim$181 test patterns, respectively.

\section{Conclusions: Key Challenges and Insights}

%Next, we discuss challenges and insights identified.
%Challenges we observed during our experiments include: % with GPT for this work.
%\textcolor{red}{this section should come after the case study section.}
%\begin{itemize}
%[noitemsep,leftmargin=*]
Designs with a large number of states are challenging for LLMs to analyze all possible transitions. 
    % %This becomes more evident with transitions where the connections are more
    % tangible.\textcolor{red}{What do you mean here, by more tangible connections?}
    Eventually, such cases lead to a repetition of responses. %Thus, LLMs shows scalability limitations.    
    Second, fine-tuning an LLM for this work's scope is challenging as testbenches depend on RTL codes. Testbenches should be accompanied
    by the corresponding RTL, and such combined datasets are not available.
    From the prior point, two further challenges follow. Even if we obtain such datasets, labeling testbenches as `good' versus `bad' in terms of coverage is not easy, at least
    not without thoroughly running the EDA tools on all designs in the first place.  Interdependency between regular RTL and testbenches may be difficult to comprehend for the LLMs, given the syntactic differences in these two types of HDL files.
    % \textcolor{red}{The latter is just my best, somewhat uninformed guess. Please revise if that's not correct. Note that I've also put some related discussion for the insights below.}

%    Thus, forcing us to use state-of-art LLM models like GPT4 to better analyze our RTL codes. 
%\end{itemize}

%\newpage
%\subsection{Insights}
% into solving these challenges:
To overcome the scalability challenge for the large numbers of states and transitions, we prompt LLMs with phrases like ``reset after X number of transitions are covered''. The model can then start with a new context to generate the remaining test patterns without getting stuck at some state. Furthermore, without access to a golden testbench for few-shot learning, the feedback from the EDA tools is important as such feedback in the prompt better guide the model. That is, after achieving some coverage, prompts are modified with phrases like ``Keep the previous test cases but now focus more on the transitions that are not covered yet''. Once we provide the underlying RTL, the LLM can follow the context. The interdependency between testbenches and RTL is better understood throughout the iterative process and evidenced through our case study on bug detection. Prompts play a crucial role in the output quality.
%of the output.

\begin{comment}
\section{Conclusion and Future Work\label{sec:conclusions}}

The framework integrates LLMs and commercial EDA tools for guided feedback to enhance the quality of testbench generation. We show the benefits of iterative, high-quality feedback using case studies on FSMs. Bugs can be detected by providing test data to the model and asking it to match them with design specifications to flag inconsistencies. These promising findings show the real-world relevance of LLM-assisted testbench generation and bug detection. We will extend our LLM-guided approach beyond FSMs to
complex modules.
%, explore LLM-guided fuzzing, and use testbenches for code repair.
\end{comment}

%\section*{Acknowledgments}
\newpage
  \newpage
   \appendix

\section{Appendix}
\subsection{Result II}
 \label{app:case}

~\autoref{tab:results_rest} reports our results for the rest of the FSMs other than the one we have mentioned in the ~\autoref{tab:results_our}. 

\begin{table*}[ht]

\caption{Remainder of the Result. See also~\autoref{tab:results_our} caption.
}
\label{tab:results_rest}
\centering
%\footnotesize
\setlength{\tabcolsep}{2.5pt}
\resizebox{\textwidth}{!}{
\begin{tabular}{|c|c|c|c|c|c|c|c|c|c|c|c|c|c|c|c|c|c|c|c|c|c|c|}
\hline
\multirow{7}{*}{ \rotatebox{90}{\textbf{Level}} } & \multirow{7}{*}{ \rotatebox{90}{\textbf{FSMs}} } & \multicolumn{3}{c|}{\multirow{3}{*}{ \textbf{Characteristics} }} & \multicolumn{6}{c|}{\textbf{LLM-guided Testbench Generation}} & \multicolumn{12}{c|}{\textbf{LLM-guided Automated Bug Detection}} \\ \cline{6-23}
& & \multicolumn{3}{c|}{} & 
{ \textbf{GPT3.5}} & 
{ \textbf{GPT4}} & \multicolumn{2}{c|}{\textbf{GPT3.5+This Work}} & \multicolumn{2}{c|}{\textbf{GPT4+This Work}} & \multicolumn{3}{c|}{ \textbf{GPT3.5} } & \multicolumn{3}{c|}
{ \textbf{GPT4} } & \multicolumn{3}{c|}{ \textbf{GPT3.5+This Work}} & \multicolumn{3}{c|}{\textbf{GPT4+This Work }} \\ 
\cline{3-23}
& & \multirow{4}{*}{ \rotatebox{90}{i/o} } & \multirow{4}{*}{ \rotatebox{90}{o/p} } & \multirow{4}{*}{ \rotatebox{90}{states} } & \multirow{4}{*}{ \rotatebox{90}{\% Cov} } & \multirow{4}{*}{ \rotatebox{90}{\% Cov} } & \multirow{4}{*}{ \rotatebox{90}{Iters} } & \multirow{4}{*}{ \rotatebox{90}{\% Cov} } & \multirow{4}{*}{ \rotatebox{90}{Iters} } & \multirow{4}{*}{ \rotatebox{90}{\% Cov}}  & \multirow{4}{*}{ \rotatebox{90}{State Regs} } & \multirow{4}{*}{ \rotatebox{90}{I/O pairs} } & \multirow{4}{*}{ \rotatebox{90}{Fuzzing} } & \multirow{4}{*}{ \rotatebox{90}{State Regs} } & \multirow{4}{*}{ \rotatebox{90}{I/O pairs} } & \multirow{4}{*}{ \rotatebox{90}{Fuzzing} } & \multirow{4}{*}{ \rotatebox{90}{State Regs} } & \multirow{4}{*}{ \rotatebox{90}{I/O pairs} } & \multirow{4}{*}{ \rotatebox{90}{Fuzzing} } & \multirow{4}{*}{ \rotatebox{90}{State Regs} } & \multirow{4}{*}{ \rotatebox{90}{I/O pairs} } & \multirow{4}{*}{ \rotatebox{90}{Fuzzing} } \\
& & & & & & & & & & & & & & & & & & & & & & \\
& & & & & & & & & & & & & & & & & & & & & & \\
& & & & & & & & & & & & & & & & & & & & & & \\ \hline
\multirow{21}{*}{ \rotatebox{90}{Easy} } 
& 3 & 1 & 2 & 2 & 50 & 50 & 2 & 100 & 2 & 100 & \checkmark & \checkmark & \checkmark & \checkmark & \checkmark & \checkmark & \checkmark & \checkmark & \checkmark & \checkmark & \checkmark & \checkmark \\
& 4 & 2 & 2 & 2 & 75 & 50 & 3 & 100 & 2 & 100 & \checkmark & \checkmark & \checkmark & \checkmark & \checkmark & \checkmark & \checkmark & \checkmark & \checkmark & \checkmark & \checkmark & \checkmark \\
& 5 & 2 & 2 & 2 & 50 & 75 & 3 & 100 & 2 & 100 & \checkmark & \checkmark & \checkmark & \checkmark & \checkmark & \checkmark & \checkmark & \checkmark & \checkmark & \checkmark & \checkmark & \checkmark \\
& 6 & 2 & 1 & 2 & 25 & 50 & 2 & 100 & 2 & 100 & \checkmark & \checkmark & \checkmark & \checkmark & \checkmark & \checkmark & \checkmark & \checkmark & \checkmark & \checkmark & \checkmark & \checkmark \\
& 7 & 2 & 2 & 2 & 50 & 75 & 2 & 100 & 2 & 100 & \checkmark & \checkmark & \checkmark & \checkmark & \checkmark & \checkmark & \checkmark & \checkmark & \checkmark & \checkmark & \checkmark & \checkmark \\
& 8 & 2 & 3 & 2 & 50 & 50 & 3 & 100 & 2 & 100 & \checkmark & \checkmark & \checkmark & \checkmark & \checkmark & \checkmark & \checkmark & \checkmark & \checkmark & \checkmark & \checkmark & \checkmark \\
& 11 & 1 & 1 & 3 & 42 & 50 & 3 & 100 & 3 & 100 & \checkmark & \checkmark & \checkmark & \checkmark & \checkmark & \checkmark & \checkmark & \checkmark & \checkmark & \checkmark & \checkmark & \checkmark \\
& 12 & 1 & 1 & 3 & 50 & 50 & 2 & 100 & 2 & 100 & \checkmark & \checkmark & \checkmark & \checkmark & \checkmark & \checkmark & \checkmark & \checkmark & \checkmark & \checkmark & \checkmark & \checkmark \\
& 13 & 3 & 3 & 3 & 24 & 50 & 4 & 95 & 2 & 100 & \checkmark & \checkmark & \checkmark & \checkmark & \checkmark & \checkmark & \checkmark & \checkmark & \checkmark & \checkmark & \checkmark & \checkmark \\
& 14 & 1 & 1 & 3 & 33 & 40 & 3 & 100 & 3 & 100 & \checkmark & \checkmark & \checkmark & \checkmark & \checkmark & \checkmark & \checkmark & \checkmark & \checkmark & \checkmark & \checkmark & \checkmark \\
& 21 & 1 & 2 & 4 & 20 & 45 & 4 & 94 & 3 & 100 & \checkmark & \checkmark & \checkmark & \checkmark & \checkmark & \checkmark & \checkmark & \checkmark & \checkmark & \checkmark & \checkmark & \checkmark \\
& 22 & 2 & 2 & 4 & 40 & 50 & 5 & 100 & 2 & 100 & \checkmark & \checkmark & \checkmark & \checkmark & \checkmark & \checkmark & \checkmark & \checkmark & \checkmark & \checkmark & \checkmark & \checkmark \\
& 23 & 2 & 2 & 4 & 25 & 25 & 5 & 100 & 2 & 100 & \checkmark & \checkmark & \checkmark & \checkmark & \checkmark & \checkmark & \checkmark & \checkmark & \checkmark & \checkmark & \checkmark & \checkmark \\
& 24 & 1 & 1 & 4 & 32 & 40 & 4 & 97 & 3 & 100 & \checkmark & \checkmark & \checkmark & \checkmark & \checkmark & \checkmark & \checkmark & \checkmark & \checkmark & \checkmark & \checkmark & \checkmark \\
& 37 & 1 & 1 & 6 & 25 & 50 & 5 & 91 & 3 & 100 & \checkmark & \checkmark & \checkmark & \checkmark & \checkmark & \checkmark & \checkmark & \checkmark & \checkmark & \checkmark & \checkmark & \checkmark \\
& 38 & 6 & 2 & 6 & 50 & 50 & 4 & 100 & 2 & 100 & \checkmark & \checkmark & \checkmark & \checkmark & \checkmark & \checkmark & \checkmark & \checkmark & \checkmark & \checkmark & \checkmark & \checkmark \\
& 39 & 3 & 3 & 6 & 50 & 45 & 6 & 100 & 2 & 100 & \checkmark & \checkmark & \checkmark & \checkmark & \checkmark & \checkmark & \checkmark & \checkmark & \checkmark & \checkmark & \checkmark & \checkmark \\
& 40 & 1 & 4 & 6 & 25 & 25 & 5 & 93 & 3 & 90 & \checkmark & \checkmark & \checkmark & \checkmark & \checkmark & \checkmark & \checkmark & \checkmark & \checkmark & \checkmark & \checkmark & \checkmark \\
& 41 & 1 & 1 & 6 & 20 & 20 & 5 & 100 & 4 & 100 & \checkmark & \checkmark & \checkmark & \checkmark & \checkmark & \checkmark & \checkmark & \checkmark & \checkmark & \checkmark & \checkmark & \checkmark \\
& 44 & 2 & 1 & 7 & 25 & 37 & 6 & 90 & 3 & 95 & \checkmark & \checkmark & \checkmark & \checkmark & \checkmark & \checkmark & \checkmark & \checkmark & \checkmark & \checkmark & \checkmark & \checkmark \\
& 45 & 1 & 4 & 7 & 50 & 50 & 7 & 100 & 4 & 100 & \checkmark & \checkmark & \checkmark & \checkmark & \checkmark & \checkmark & \checkmark & \checkmark & \checkmark & \checkmark & \checkmark & \checkmark \\ \hline
\multirow{19}{*}{ \rotatebox{90}{Medium} } 
& 49 & 1 & 2 & 8 & 25 & 30 & 6 & 90 & 5 & 100 & \checkmark & \checkmark & \checkmark & \checkmark & \checkmark & \checkmark & \checkmark & \checkmark & \checkmark & \checkmark & \checkmark & \checkmark \\
& 50 & 2 & 2 & 8 & 20 & 35 & 6 & 95 & 4 & 100 & \checkmark & \checkmark & \checkmark & \checkmark & \checkmark & \checkmark & \checkmark & \checkmark & \checkmark & \checkmark & \checkmark & \checkmark \\
& 51 & 1 & 4 & 8 & 15 & 40 & 7 & 91 & 4 & 100 & \checkmark & \checkmark & \checkmark & \checkmark & \checkmark & \checkmark & \checkmark & \checkmark & \checkmark & \checkmark & \checkmark & \checkmark \\
& 55 & 2 & 1 & 9 & 25 & 38 & 7 & 93 & 5 & 100 & \checkmark & \checkmark & \checkmark & \checkmark & \checkmark & \checkmark & \checkmark & \checkmark & \checkmark & \checkmark & \checkmark & \checkmark \\
& 56 & 3 & 3 & 9 & 18 & 25 & 6 & 90 & 4 & 95 & \checkmark & \checkmark & \checkmark & \checkmark & \checkmark & \checkmark & \checkmark & \checkmark & \checkmark & \checkmark & \checkmark & \checkmark \\
& 57 & 1 & 4 & 9 & 33 & 25 & 7 & 90 & 6 & 91 & \checkmark & \xmark & \checkmark & \checkmark & \checkmark & \checkmark & \checkmark & \checkmark & \checkmark & \checkmark & \checkmark & \checkmark \\
& 58 & 1 & 1 & 9 & 12 & 18 & 5 & 94 & 5 & 95 & \checkmark & \xmark & \checkmark & \checkmark & \checkmark & \checkmark & \checkmark & \checkmark & \checkmark & \checkmark & \checkmark & \checkmark \\
& 63 & 5 & 5 & 10 & 15 & 10 & 9 & 96 & 7 & 100 & \checkmark & \xmark & \xmark & \checkmark & \checkmark & \checkmark & \checkmark & \checkmark & \checkmark & \checkmark & \checkmark & \checkmark \\
& 64 & 1 & 1 & 10 & 20 & 45 & 8 & 90 & 5 & 100 & \xmark & \xmark & \xmark & \checkmark & \checkmark & \checkmark & \checkmark & \checkmark & \checkmark & \checkmark & \checkmark & \checkmark \\
& 65 & 1 & 2 & 10 & 18 & 25 & 8 & 100 & 5 & 93 & \xmark & \xmark & \xmark & \checkmark & \xmark & \checkmark & \checkmark & \checkmark & \checkmark & \checkmark & \checkmark & \checkmark \\
& 68 & 2 & 1 & 11 & 10 & 15 & 9 & 100 & 6 & 96 & \xmark & \xmark & \xmark & \checkmark & \xmark & \checkmark & \checkmark & \checkmark & \checkmark & \checkmark & \checkmark & \checkmark \\
& 69 & 3 & 1 & 11 & 20 & 20 & 9 & 92 & 7 & 90 & \xmark & \xmark & \xmark & \checkmark & \xmark & \checkmark & \checkmark & \checkmark & \checkmark & \checkmark & \checkmark & \checkmark \\
& 70 & 1 & 1 & 11 & 25 & 20 & 8 & 98 & 6 & 100 & \xmark & \xmark & \xmark & \checkmark & \xmark & \checkmark & \checkmark & \checkmark & \checkmark & \checkmark & \checkmark & \checkmark \\
& 73 & 2 & 2 & 12 & 35 & 41 & 11 & 90 & 8 & 90 & \xmark & \xmark & \xmark & \checkmark & \xmark & \xmark & \checkmark & \checkmark & \checkmark & \checkmark & \checkmark & \checkmark \\
& 74 & 3 & 1 & 12 & 17 & 15 & 10 & 95 & 7 & 95 & \xmark & \xmark & \xmark & \xmark & \xmark & \xmark & \checkmark & \checkmark & \checkmark & \checkmark & \checkmark & \checkmark \\
& 76 & 4 & 3 & 13 & 9 & 18 & 11 & 90 & 9 & 90 & \xmark & \xmark & \xmark & \xmark & \xmark & \xmark & \checkmark & \checkmark & \checkmark & \checkmark & \checkmark & \checkmark \\
& 77 & 3 & 3 & 13 & 11 & 10 & 10 & 75 & 8 & 90 & \xmark & \xmark & \xmark & \xmark & \xmark & \xmark & \checkmark & \checkmark & \checkmark & \checkmark & \checkmark & \checkmark \\
& 79 & 1 & 1 & 14 & 25 & 22 & 11 & 91 & 9 & 92 & \xmark & \xmark & \xmark & \xmark & \xmark & \xmark & \checkmark & \checkmark & \checkmark & \checkmark & \checkmark & \checkmark \\
& 80 & 1 & 2 & 14 & 32 & 10 & 12 & 90 & 9 & 100 & \xmark & \xmark & \xmark & \xmark & \xmark & \xmark & \checkmark & \checkmark & \checkmark & \checkmark & \checkmark & \checkmark \\ \hline
\multirow{10}{*}{ \rotatebox{90}{Hard }} 
& 81 & 1 & 1 & 15 & 13 & 16 & 14 & 100 & 10 & 95 & \xmark & \xmark & \xmark & \xmark & \xmark & \xmark & \checkmark & \checkmark & \checkmark & \checkmark & \checkmark & \checkmark \\
& 83 & 2 & 2 & 17 & 11 & 17 & 16 & 90 & 12 & 96 & \xmark & \xmark & \xmark & \xmark & \xmark & \xmark & \checkmark & \xmark & \checkmark & \checkmark & \checkmark & \checkmark \\
& 84 & 1 & 1 & 18 & 6 & 8 & 19 & 95 & 13 & 97 & \xmark & \xmark & \xmark & \xmark & \xmark & \xmark & \checkmark & \xmark & \xmark & \checkmark & \checkmark & \checkmark \\
& 86 & 1 & 2 & 19 & 8 & 5 & 20 & 92 & 13 & 91 & \xmark & \xmark & \xmark & \xmark & \xmark & \xmark & \xmark & \xmark & \xmark & \checkmark & \checkmark & \checkmark \\
& 88 & 1 & 1 & 20 & 9 & 6 & 21 & 91 & 14 & 92 & \xmark & \xmark & \xmark & \xmark & \xmark & \xmark & \xmark & \xmark & \xmark & \checkmark & \checkmark & \checkmark \\
& 89 & 1 & 1 & 21 & 5 & 9 & 21 & 93 & 12 & 95 & \xmark & \xmark & \xmark & \xmark & \xmark & \xmark & \xmark & \xmark & \xmark & \checkmark & \checkmark & \checkmark \\
& 91 & 3 & 3 & 22 & 10 & 11 & 23 & 90 & 12 & 90 & \xmark & \xmark & \xmark & \xmark & \xmark & \xmark & \xmark & \xmark & \xmark & \checkmark & \checkmark & \checkmark \\
& 94 & 1 & 1 & 24 & 4 & 13 & 24 & 96 & 13 & 100 & \xmark & \xmark & \xmark & \xmark & \xmark & \xmark & \xmark & \xmark & \xmark & \checkmark & \checkmark & \checkmark \\
& 97 & 1 & 1 & 26 & 8 & 6 & 23 & 90 & 14 & 90 & \xmark & \xmark & \xmark & \xmark & \xmark & \xmark & \xmark & \xmark & \xmark & \checkmark & \checkmark & \checkmark \\
& 98 & 1 & 1 & 26 & 7 & 3 & 22 & 90 & 14 & 95 & \xmark & \xmark & \xmark & \xmark & \xmark & \xmark & \xmark & \xmark & \xmark & \checkmark & \checkmark & \checkmark \\
\hline
\end{tabular}
}
\end{table*}

\newpage

\bibliographystyle{IEEEtran}
%\scriptsize{
\bibliography{biblio.bib}

% Generated by IEEEtran.bst, version: 1.12 (2007/01/11)
\begin{thebibliography}{10}
\providecommand{\url}[1]{#1}
\csname url@samestyle\endcsname
\providecommand{\newblock}{\relax}
\providecommand{\bibinfo}[2]{#2}
\providecommand{\BIBentrySTDinterwordspacing}{\spaceskip=0pt\relax}
\providecommand{\BIBentryALTinterwordstretchfactor}{4}
\providecommand{\BIBentryALTinterwordspacing}{\spaceskip=\fontdimen2\font plus
\BIBentryALTinterwordstretchfactor\fontdimen3\font minus \fontdimen4\font\relax}
\providecommand{\BIBforeignlanguage}[2]{{%
\expandafter\ifx\csname l@#1\endcsname\relax
\typeout{** WARNING: IEEEtran.bst: No hyphenation pattern has been}%
\typeout{** loaded for the language `#1'. Using the pattern for}%
\typeout{** the default language instead.}%
\else
\language=\csname l@#1\endcsname
\fi
#2}}
\providecommand{\BIBdecl}{\relax}
\BIBdecl

\bibitem{takahashi03}
J.~Takahashi \emph{et~al.}, ``Testing for high assurance system by fsm,'' \emph{IEICE TRANSACTIONS on Information and Systems}, vol.~86, no.~10, pp. 2114--2120, 2003.

\bibitem{openai2024gpt4}
\BIBentryALTinterwordspacing
OpenAI, ``\BIBforeignlanguage{en-US}{{GPT}-4},'' Mar. 2023. [Online]. Available: \url{https://openai.com/research/gpt-4}
\BIBentrySTDinterwordspacing

\bibitem{liu2023verilogeval}
M.~Liu \emph{et~al.}, ``Verilogeval: Evaluating large language models for verilog code generation,'' in \emph{2023 IEEE/ACM International Conference on Computer Aided Design (ICCAD)}.\hskip 1em plus 0.5em minus 0.4em\relax IEEE, 2023, pp. 1--8.

\bibitem{thakur2023autochip}
S.~Thakur \emph{et~al.}, ``Autochip: Automating hdl generation using llm feedback,'' \emph{arXiv preprint arXiv:2311.04887}, 2023.

\bibitem{lu2023rtllm}
Y.~Lu \emph{et~al.}, ``Rtllm: An open-source benchmark for design rtl generation with large language model,'' in \emph{2024 29th Asia and South Pacific Design Automation Conference (ASP-DAC)}.\hskip 1em plus 0.5em minus 0.4em\relax IEEE, 2024, pp. 722--727.

\bibitem{thakur2023verigen}
S.~Thakur \emph{et~al.}, ``Verigen: A large language model for verilog code generation,'' \emph{ACM Transactions on Design Automation of Electronic Systems}, 2023.

\bibitem{kande2023llmassisted}
R.~Kande \emph{et~al.}, ``Llm-assisted generation of hardware assertions,'' \emph{arXiv preprint arXiv:2306.14027}, 2023.

\bibitem{fang2024assertllm}
W.~Fang \emph{et~al.}, ``Assertllm: Generating and evaluating hardware verification assertions from design specifications via multi-llms,'' \emph{arXiv preprint arXiv:2402.00386}, 2024.

\bibitem{wu2024chateda}
H.~Wu \emph{et~al.}, ``Chateda: A large language model powered autonomous agent for eda,'' \emph{IEEE Transactions on Computer-Aided Design of Integrated Circuits and Systems}, 2024.

\bibitem{liu2023chipnemo}
M.~Liu \emph{et~al.}, ``Chipnemo: Domain-adapted llms for chip design,'' \emph{arXiv preprint arXiv:2311.00176}, 2023.

\bibitem{zhong2023llm4eda}
R.~Zhong \emph{et~al.}, ``Llm4eda: Emerging progress in large language models for electronic design automation,'' \emph{arXiv preprint arXiv:2401.12224}, 2023.

\bibitem{Blocklove_2023}
J.~Blocklove \emph{et~al.}, ``Chip-chat: Challenges and opportunities in conversational hardware design,'' in \emph{2023 ACM/IEEE 5th Workshop on Machine Learning for CAD (MLCAD)}.\hskip 1em plus 0.5em minus 0.4em\relax IEEE, 2023, pp. 1--6.

\bibitem{fu2023gpt4aigchip}
Y.~Fu \emph{et~al.}, ``Gpt4aigchip: Towards next-generation ai accelerator design automation via large language models,'' in \emph{2023 IEEE/ACM International Conference on Computer Aided Design (ICCAD)}.\hskip 1em plus 0.5em minus 0.4em\relax IEEE, 2023, pp. 1--9.

\bibitem{chang2023chipgpt}
K.~Chang \emph{et~al.}, ``Chipgpt: How far are we from natural language hardware design,'' \emph{arXiv preprint arXiv:2305.14019}, 2023.

\bibitem{hdlbits}
\BIBentryALTinterwordspacing
``Problem sets - {HDLBits},'' 2023. [Online]. Available: \url{https://hdlbits.01xz.net/wiki/Problem_sets}
\BIBentrySTDinterwordspacing

\bibitem{github}
\BIBentryALTinterwordspacing
J.~Bhandari. (2024) {LLM-Aided-Testbench-Generation-for-FSM}. [Online]. Available: \url{https://github.com/jitendra-bhandari/LLM-Aided-Testbench-Generation-for-FSM}
\BIBentrySTDinterwordspacing

\end{thebibliography}
%}

% \bibliographystyle{IEEEtran}
% \bibliography{biblio}

\end{document}